\documentclass[prodmode,acmtist]{acmsmall} % Aptara syntax

\usepackage{amsmath}
\usepackage{natbib}
\usepackage[tableposition=top]{caption}
\usepackage{tabularx}
\usepackage{algorithmicx}
\usepackage{algpseudocode}
\usepackage{url}
\acmVolume{x}
\acmNumber{x}
\acmArticle{xx}
\acmYear{2018}
\acmMonth{10}

\begin{document}

% Page heads
\markboth{B. Loni et al.}{Factorization Machines for Data with Implicit Feedback}

% Title portion
\title{Factorization Machines for Datasets with Implicit Feedback}
\author{Babak Loni
\affil{Delft University of Technology}
Martha Larson
\affil{Delft University of Technology}
Alan Hanjalic
\affil{Delft University of Technology}
}
% NOTE! Affiliations placed here should be for the institution where the
%       BULK of the research was done. If the author has gone to a new
%       institution, before publication, the (above) affiliation should NOT be changed.
%       The authors 'current' address may be given in the "Author's addresses:" block (below).
%       So for example, Mr. Abdelzaher, the bulk of the research was done at UIUC, and he is
%       currently affiliated with NASA.

\begin{abstract}
Factorization Machines (FMs) are generic factorization models for Collaborative Filtering (CF) that offer several advantages compared to the conventional CF models. 
They are expressive and general models that can mimic several CF problems, their accuracy is state-of-the-art, and their linear complexity makes them fast and scalable.
Factorization Machines however, are optimized for datasets with explicit feedback (such as ratings) and they are not very effective for datasets with \textit{implicit} feedback.
%Existing research on FMs for implicit feedback datasets are limited or have not addressed the problem thoroughly.
%This can be due to the fact that the optimization model in FMs is a \textit{point-wise} method with a squared-error loss function that is not a proper optimization model for implicit feedback. 
%A trivial way to use FMs for datasets with implicit feedback is to map user's feedback to real-value numbers and train the model with the standard FMs model.
Although FMs can also be used for datasets with implicit feedback by a trivial mapping of implicit feedback to explicit values, but we will empirically show that such trivial mapping is not optimized for ranking.
%In this work we first show that the trivial mapping is not an effective way to use FMs for datasets with implicit feedback. 
In this work, we propose \textit{FM-Pair}, an adaptation of Factorization Machines with a \textit{pairwise} loss function, making them effective for datasets with implicit feedback.
The optimization model in FM-Pair is based on the BPR (Bayesian Personalized Ranking) criterion, which is a well-established pairwise optimization model.
FM-Pair retains the advantages of FMs on generality, expressiveness and performance and yet it can be used for datasets with implicit feedback.
%The main advantage of FM-Pair compared to other learning-to-rank methods such as BPR-MF is that it can exploit any auxiliary information just like the standard FMs model by adding extra features.
We also propose how to apply FM-Pair effectively on two collaborative filtering problems, namely, context-aware recommendation and cross-domain collaborative filtering. 
By performing experiments on different datasets with explicit or implicit feedback we empirically show that in most of the tested datasets, FM-Pair beats state-of-the-art learning-to-rank methods such as BPR-MF (BPR with Matrix Factorization model). We also show that FM-Pair is significantly more effective for ranking, compared to the standard FMs model. Moreover, we show that FM-Pair can utilize context or cross-domain information effectively as the accuracy of recommendations would always improve with the right auxiliary features. Finally we show that FM-Pair has a linear time complexity and scales linearly by exploiting additional features.

\end{abstract}




\begin{bottomstuff}
This work is supported by the EU FP7 project Crowdrec, under
grant agreements no. 610594.

Author's addresses: Delft University of Technology, Mekelweg 4, 2628CD Delft, The Netherlands.
\end{bottomstuff}

\maketitle

\section{Introduction}\label{sec:Introduction}
The role of Recommender Systems (RS) as a tool to provide personalized content for users is becoming more and more important. Among different techniques that have been introduced for recommender systems in the past two decades, Collaborative Filtering (CF) has become the most successful and widely-used technique. Early CF techniques were mainly based on neighborhood approaches \cite{Resnick94} while recently model-based techniques such as Matrix Factorization (MF) \cite{salakhutdinov08} attracted more attention due to their superior performance and scalability \cite{Koren09}.
Matrix factorization techniques generally learn a low-dimensional representation of users and items by mapping them into a joint latent space consisting of latent \emph{factors}. Recommendations are then generated based on the similarity of user and item factors.

Factorization Machines (FMs) \cite{Rendle10} are general factorization models that not only learn user and item latent factors, but also the relation between users and items with any \textit{auxiliary}\footnote{We use the term ``auxiliary" for any additional information that is available next to the user-item matrix. Auxiliary features can be user and item attributes, context, content, information from other domains and so on.} features. This is done by also factorizing auxiliary features to the same joint latent space.
%fMs are similar to regression models, but adapted for problems with high dimensional data specifically for collaborative filtering. 
In contrast to the conventional factorization techniques where the training data is represented by a matrix, the input data for FMs are feature vectors just similar to the input data for the other supervised learning methods such as Support Vector Machines or regression models. This creates a great flexibility for FMs by allowing them to incorporate any additional information in terms of auxiliary features. Thanks to the flexibility of FMs on representing the data, FMs can mimic other factorization models by feature engineering without the need to change the underlying model. In~\cite{Rendle12} Rendle shows how several factorization model such as Matrix Factorization~\cite{Koren09}, Attribute-Aware models~\cite{Gantner10} and SVD++~\cite{Koren08} can be mimic by FMs. 
Factorization Machines have been successfully applied in different collaborative filtering problems including context-aware recommendation~\cite{Rendle11_2}, cross-domain collaborative filtering~\cite{Loni2014} and social recommendation~\cite{Zhang13}. Factorization Machines have linear time complexity and thus are scalable for large datasets. They have been shown~\cite{Rendle11_2} to be significantly faster than tensor factorization~\cite{Karatzoglou10}, a popular context-aware CF method. Moreover, an effective parallel optimization method for FMs has been developed~\cite{Juan16}, reporting significant speed-up in training time compared to the standard training models.

Despite the great advantages of Factorization Machines, the FMs model is not optimized for data with implicit user feedback . All the aforementioned studies have been developed for datasets with explicit feedback. 
Implicit user feedback (such as clicks) are typically unary or positive-only feedback, and thus there are no explicit real-valued scores for user-item interactions. 
A trivial approach to train FMs with such datasets, is to \textit{map} positive feedback to a real-value number such as +1. Negative examples can be sampled from unobserved interactions and can be assigned a real-valued label of 0 or -1. The model can then be trained just like the standard FMs model.
However, such mapping methods are associated with two problems: firstly, the sampled interactions with negative labels might not be a real negative feedback as the user might not have the chance to observe the item~\cite{Rendle09}. Secondly, due to the \textit{point-wise} optimization techniques in FMs, the model learns to correctly predict +1s and -1s, which is not necessarily the optimal model for ranking. We experimentally show that such trivial mapping is not an accurate model for learning from implicit feedback.

The existing work on FMs with implicit feedback is limited and the scope of existing experimental studies is narrow. In a recent work, \citet{Guo16} introduce PRFM (Pairwise Ranking Factorization Machines), where they adapt a pairwise optimization technique to learn from implicit feedback. This work however, has not fully exploited one of the main advantages of FMs, namely the ability to encode additional information as auxiliary features. In PRFM, contextual information has been exploited to adapt the \textit{sampling} stage of the learning method and thus the model needs to be re-adapted for different types of auxiliary information. Furthermore, PRFM was only tested on one explicit dataset. In \citet{Guo16},  explicit feedback was mapped to unary positive-only feedback, and thus it is not clear how PRFM performs on datasets with inherent implicit feedback. In another work, \citet{Nguyen14} introduced Gaussian Process Factorization Machines (GPFM), a \textit{non-linear} adaptation of Factorization Machines. In GPFM, interactions between users, items and context are captured with non-linear Gaussian kernels. They also introduced a pairwise optimization model for GPFM for datasets with implicit feedback and used it for context-aware recommendations. However, GPFM is not linearly scalable as the underlying optimization method relies on calculating the inverse of Gaussian kernels, which is a computationally-intensive task. Furthermore, the GPFM model is dependent on the choice of kernel, thus making it more complex compared to the standard FMs model.
Nevertheless, we empirically compare the performance of GPFM with our method based on the training time and recommendations accuracy, in Section \ref{sec:experiments}.
%However, with increasing number of context dimensions, the task of context-aware recommendation become harder~\cite{hidasi2016general}. 

%In contrast to the point-wise optimization model of FMs, the model can be learned with a \textit{pair-wise} optimization method such as Bayesian Personalized Ranking (BPR)~\cite{Rendle09}. With pair-wise optimization techniques, the model learns to correctly \textit{rank} a given pair of items with respect to a user. Therefore, the model is optimized for ranking (not prediction) and there is no need for explicit mapping as the optimization technique just need to know the correct order of item pairs. 

In this work we consolidate previous work on FMs and introduce a generic Factorization Machines framework for datasets with implicit feedback. Similar to \cite{Guo16}, we adapt a pairwise optimization method based on BPR (Bayesian Personalized Ranking) criterion~\cite{Rendle09}. BPR is an state-of-the-art learning-to-rank method for collaborative filtering that learns to correctly rank \textit{pairs} of items with respect to a user. BPR has been successfully applied for datasets with implicit feedback and it has been shown~\cite{Rendle09} to be more effective than other learning-to-rank methods such as Weighted Regularized Matrix Factorization (WRMF)~\cite{Hu08}. 
%Pair-wise models would be ideal optimization techniques for FMs when the input data is implicit feedback. 
Unlike \cite{Guo16}, our proposed training model does not depend on the sampling method. We adapt the optimization model of BPR based on the existing auxiliary information that are represented as additional features. 
We refer to our implementation as \textit{FM-Pair} since we are using a pair-wise optimization method for FMs. 
FM-Pair is a linear model that can exploit additional information as auxiliary features just like the standard FMs, without requiring any adaptation to the underlying model. 
We further propose two applications of FM-Pair on context-aware and cross-domain collaborative filtering problems.
We test FM-Pair on four implicit and explicit datasets with different tasks. 
We find that by using the right data, FM-Pair outperforms state-of-the-art methods for learning from implicit feedback data. We also show that FM-Pair is significantly more effective than the trivial implicit-to-explicit mapping method.
Furthermore, we empirically demonstrate the effectiveness of FM-Pair on exploiting auxiliary features (i.e., context or cross-domain information). We also empirically show that FM-Pair scales linearly by increasing dimensionality of factorization or number of auxiliary features.

FM-Pair is publicly available as a part of WrapRec\footnote{http://wraprec.crowdrec.eu}, an open-source evaluation framework for recommender systems and can be used simply on different datasets.

The contributions of this work can be summarized as follows: 
\begin{itemize}
	\item An extension to Factorization Machines is introduced that allows the use of FMs for datasets with implicit feedback. Inspired by BPR~ \cite{Rendle09}, FM-Pair is implemented with a pairwise loss function without requiring explicit feedback for user-item interactions. 
	
	\item We propose to apply FM-Pair to exploit context and provide context-aware recommendations for datasets with implicit feedback. Similarly, we propose a method to apply FM-Pair for the task of cross-domain collaborative filtering.
	
	\item We release the implementation of FM-Pair as a part of WrapRec, an evaluation framework for recommender systems. The usage of WrapRec framework is briefly described in this work.
\end{itemize}

%In this work we are going to answer the following research questions:
%\begin{itemize}
%	\item RQ0: As a sanity check, is a naive implicit to explicit feedback mapping a proper way to use factorization machines when we only have implicit feedback.
	
%	\item RQ1: Can we use Factorization Machines for implicit feedback and still benefit from auxiliary features just like the standard FMs. 
	
%	\item RQ2: How can we use FMs with implicit feedback for context-aware and cross-domain CF problems.
%\end{itemize}

In the remainder of this paper we first provide a brief introduction to FMs and discuss some background and related work. 
In Section 3, we introduce FM-Pair and its pairwise optimization method in detail.
In Section 4, we propose two applications of FM-Pair for context-aware and cross-domain collaborative filtering.
In Section 5 we describe the datasets, our evaluation method and the experiments that we performed in this work and further elaborate on the results of those experiments.
We conclude the paper by summarizing the contributions and discussing about possible extensions in Section 6. 

\section{Background and Related Work}
In this section we briefly introduce the model of Factorization Machines for explicit feedback datasets and explain how the input data is represented in FMs. We also review the related work on Factorization Machines in more details.

\begin{figure}[b]
	\centering
	\includegraphics[width=0.9\textwidth]{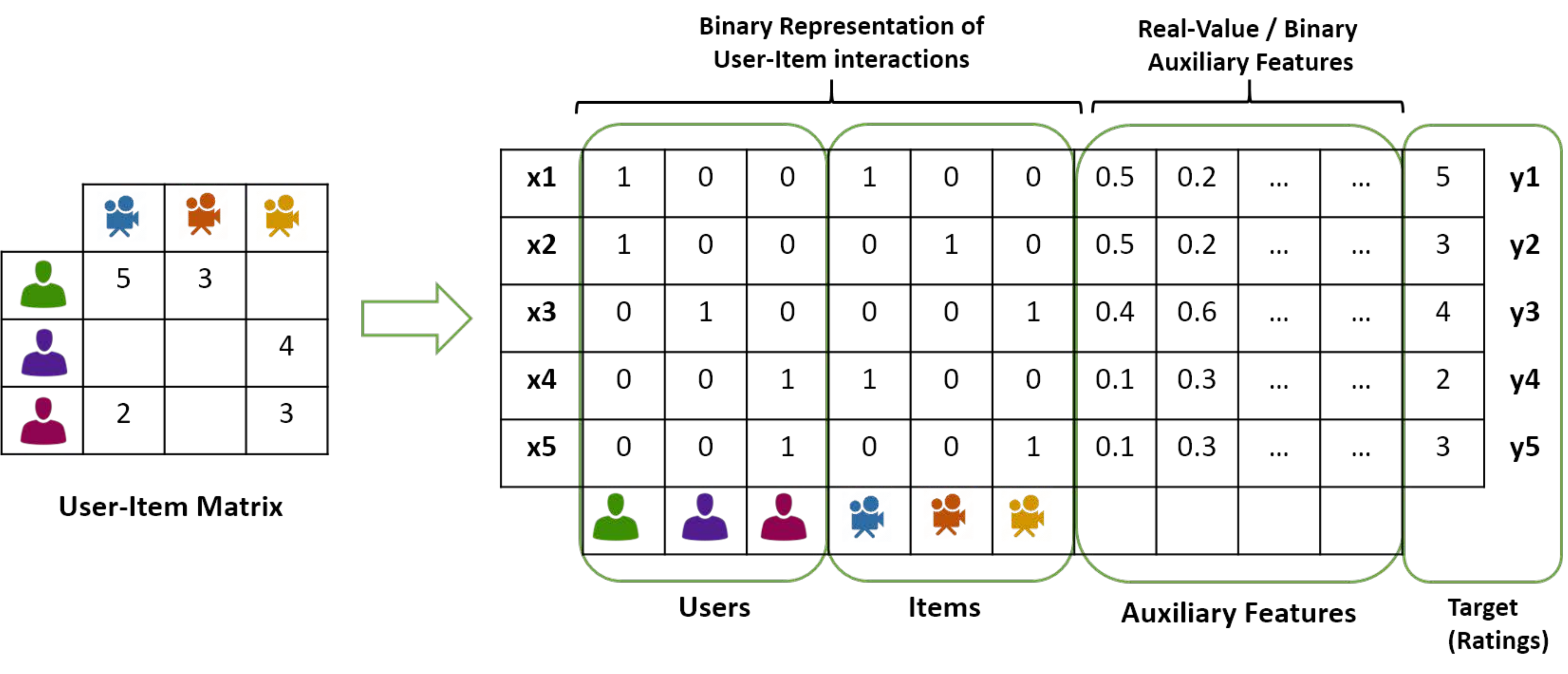}
	\caption{An overview of data representation in Factorization Machines.}
	\label{fig:overview}
\end{figure}

%\subsection{Factorization Machines for Explicit Feedback}
Factorization Machines represent each user-item interaction by a real-valued feature vector $\mathbf{x}$ with a corresponding output value of $y$. Typically the feature vectors are binary vectors with two non-zero features corresponding to user and item. In case of explicit user feedback, the output value $y$ would be the actual rating given by the user. 
%If additional information are present, they can be represented by auxiliary features.
%For each user and item one feature is considered by which we can make a binary representation indicating which user rated which item. The feature vectors can be optionally extended by auxiliary feature, if available. 
Figure \ref{fig:overview} illustrates how the user-item rating matrix can be modeled by the feature vectors $\mathbf{x}$ and output values $y$. Each rating in the user-item matrix is represented by a feature vector. The feature vectors indicate which user rated which item and if auxiliary information (such as context) is available for the user-item interactions, it is represented by real-valued auxiliary features.

More specifically, let us assume that the input data is represented by a set $S$ of tuples $(\mathbf{x},y)$ where $\mathbf{x} = (x_1,\dots,x_n) \in \mathbb{R}^n$ is a $n$-dimensional
feature vector and $y$ is its corresponding output value. 
Factorization Machines learn a model based on the interaction between features. The FM model with the order of 2, where the interactions up to order of 2 (i.e., pairs) are considered, is represented as follows:
%Factorization machines model all interactions between features using factorized interaction parameters. In this work we adopted a FM model with order $d=2$ where only \emph{pairwise} interaction between features are considered. This model can be represented as follows:
\begin{equation}
\label{eq:fm}
f(\mathbf{x}) = w_0 + \sum_{j=1}^{n}{w_j x_j} + \sum_{j=1}^{n}{\sum_{j'=j+1}^{n} w_{j,j'} x_j x_{j'}}
\end{equation}
where $w_j$ are first order interaction parameters and $w_{j,j'}$ are second order factorized interaction parameters and are defined as $w_{j,j'} = \langle \mathbf{v}_j
. \mathbf{v}_{j'} \rangle$ where $\mathbf{v}_j=(v_{j,1}, \dots, v_{j,k})$ is $k$-dimensional \textit{factorized} vector for feature $j$. In fact, FMs factorize any feature that is represented in a feature vector $\textbf{x}$ and consider the terms $x_j x_j'$ as weight parameters for the pairwise interactions between the factorized parameters.
As you might notice, the FM model is similar to a polynomial regression model. However,  FMs differ from a standard polynomial regression model by the fact that the parameters $w_{j,j'}$ are not independent parameters as they are inner product of two factorized vectors. This makes the total number of parameters much lower (compared to the regression model) and makes FMs favorable for problems with sparse and high dimensional data such as collaborative filtering.
For a FM with $n$ as the dimensionality of feature vectors and $k$ as the dimensionality of factorization, the model parameters that need to be learned are
$\Theta = \{w_0,w_1,\dots,w_n,v_{1,1},\dots,v_{n,k}\}$. 

\citet{Rendle12} proposes three learning methods to learn the parameters of FMs: Stochastic Gradient Descent (SGD), Alternating Least-Squares (ALS) and Markov Chain Monte Carlo (MCMC) method. In principal all the three methods find the optimal parameters by optimizing the same objective function but they use different techniques to solve the optimization problem. The objective function is defined by summing up the losses of individual samples in the training set. A regularization term is also added to the objective function to prevent over-fitting. The objective function $L$ with square loss over training set $S$ is defined as:
\begin{equation}
\label{eq:loss}
L(\Theta, S) = \sum_{(\mathbf{x},y) \in S} (f(\mathbf{x}|\Theta)-y)^2 + \sum_{\theta \in \Theta} \lambda_{\theta} \theta^2
\end{equation}
where $\theta \in \Theta$ are model parameters and $\lambda_{\theta}$ is regularization value for parameter $\theta$. The optimal parameters $\Theta_{OPT}$ are found by minimizing the objective function, i.e., $\Theta_{OPT} = \text{argmin}_{\Theta} L(\Theta, S)$. \citet{Rendle12} showed that all three learning methods has the same time complexity. The advantage of MCMC over the other two optimization techniques is that it is insensitive to hyper-parameters (such as regularization values) which can avoid time-consuming search for hyper-parameters. On the other the advantages of the SGD technique are its simplicity and lower storage complexity. Details about the optimization techniques can be found in \citet{Rendle12}. 
%The above loss function is a \textit{point-wise} function, that is, the total loss is calculated based on the loss of single \textit{points} (samples) in the training set. 

Factorization Machines have several advantages compared to other factorization methods:
\begin{itemize}
	\item \textbf{Generalization}: Factorization Machines are \textit{general} factorization models. Despite other factorization models (such as matrix factorization) where specific entities (e.g. user, item) are factorized, in FMs \textit{any} dimension that can be represented in terms of a feature can be factorized to a low-dimensional latent space. In matrix factorization, predictions are generated by taking into account the interaction between user and item, but in FMs the  predictions are generated by taking into account the pairwise interaction between \textit{any} pair of features (including user and item). FMs can even take into account higher order interactions between features. 
	
	\item \textbf{Expressiveness}: The fact that the input data in FMs are represented by feature vectors, not only makes FMs easy to use but also makes it possible to mimic several collaborative filtering methods such as Tensor Factorization~\cite{Karatzoglou10} by feature engineering. This obviate the need to introduce a new prediction and inference methods for such cases. Other example of CF methods that can be represented by FMs are SVD++~\cite{Koren08}, Attribute-Aware matrix factorization~\cite{Gantner10} and Joint Matrix Factorization~\cite{Shi13}. 
	
	\item \textbf{Performance and Scalability}: The complexity of prediction and inference in FMs are linear in terms of number of latent factors and the number of non-zero features~\cite{Rendle12} and thus FMs can be scaled for large datasets. Furthermore, a parallel implementation of FMs with shared-memory architecture has been proposed~\cite{Juan16} that can achieve noticeable training speed-up on large datasets.
		
\end{itemize}

\section{Learning Factorization Machines from Implicit Feedback}

In this section we introduce FM-Pair, an adaptation of Factorization Machines with a pairwise optimization methods. Previous studies~\cite{Rendle09,Shi14,Nguyen14} have reported better performance for pairwise learning-to-rank methods compared to point-wise methods for datasets with implicit feedback. While FM-Pair benefits from the effectiveness of pairwise learning-to-rank, it can also leverage the arbitrary auxiliary features that might be available in the input data. 

The optimization technique in FM-Pair is inspired by the BPR~\cite{Rendle09} optimization criterion. The BPR method comes with an assumption that all observed positive feedback is preferred over the missing preferences. The training data in BPR consist of a user and a pair of items where the first item, referred as the \textit{positive item}, is chosen from the user positive feedback and the second item, referred as the \textit{negative item}, is sampled from the unobserved interactions. 
More specifically, the training data in BPR would be a set of tuples $S_P=\{(u,i,j)|i \in I^+_u \land j \in I \backslash I^+_u \}$ where $I^+_u$ is set of all positive feedback from user $u$, $i$ is an item with positive feedback from $u$ and $j$ is a missing item for that user which is sampled uniformly from the unobserved items. 
The BPR optimization technique learns to correctly rank the items in any given pair of items, with respect to a given user.

%note about sampling can be added here
%Since number of such tuples are quadratic in number of items, typically a sampling is done to keep the size of training set reasonable. BPR originally samples unobserved items uniformly, however non-uniform sampling methods have also been proposed for BPR \cite{Gantner12,Rendle14} which can improve performance of the trained model if item popularity has a tailed distribution. In this work we stick to the uniform sampling method since sampling is not the focus of this work.

In FM-Pair arbitrary information can be represented by auxiliary features and thus the pairwise learning method should be able to exploit those features. Let us assume that $\mathbf{z}(u,i)$ are the auxiliary features associated with user $u$ and item $i$. Then the tuple $(u,i,j) \in S_P$ indicates user $u$ prefers item $i$ over item $j$ under the observed auxiliary features $\mathbf{z}(u,i)$. Auxiliary features can be user features, item features, context or additional information about user-item interaction. FM-Pair finds the optimal parameters $\Theta$ by maximizing the following likelihood function:
\begin{equation}
\label{eq:bpr-ll}
\prod_{(u,i,j) \in S_P} p(i >_{u,\mathbf{z}} j|\Theta) 
\end{equation}
where $i >_{u,\mathbf{z}} j$ indicates item $i$ is preferred over $j$ by user $u$ under auxiliary features $\mathbf{z}=\mathbf{z}(u,i)$. Similar to the BPR model, the probability $p(i >_{u,\mathbf{z}} j|\Theta)$ is defined by mapping a utility function $g_{\mathbf{z}}(u,i,j)$ to a value between 0 and 1. This can be done by the sigmoid function $\sigma(x)=\frac{1}{1+e^{-x}}$. Therefore:
\begin{equation}
p(i >_{u,\mathbf{z}} j|\Theta) = \sigma(g_{\mathbf{z}}(u,i,j|\Theta))
\end{equation}
\vspace{1mm}

The utility function $g$ captures the interaction between user $u$, item $i$ and item $j$ with presence of auxiliary features $\mathbf{z}(u,i)$. Similar to BPR, the utility function $g$ is defined by calculating the difference between the utility of individual interactions. FM-Pair calculates the utility of individual interactions by taking into account the auxiliary features $\mathbf{z}(u,i)$. We define the utility function $g$ as:
\begin{equation}
\label{eq:utility-g}
g_{\mathbf{z}}(u,i,j|\Theta) = f_{\mathbf{z}}(u,i|\Theta) - f_{\mathbf{z}}(u,j|\Theta)
\end{equation}

The utility of individual interactions $f_{\mathbf{z}}(u,i|\Theta)$ can be calculated using equation \eqref{eq:fm}. 
In this case the input feature vector $\mathbf{x}$ is a sparse vector in which features corresponding to user $u$, item $i$ and $\mathbf{z}(u,i)$ are non-zero. Thus the vector $\mathbf{x}$ can be represented with the following sparse form:
\begin{equation}
\mathbf{x}(u,i,\mathbf{z}) = \mathbf{x}_{u,i,\mathbf{z}} = \{(u,x_u),(i,x_z),\{(z,x_z)|z \in \mathbf{z}\}\}
\end{equation}
where $x_z$ is the value of feature $z$ and can be a real value number. The parameters $x_u$ and $x_i$  are considered to be 1 to indicate the corresponding user and item of a feedback (see Figure \ref{fig:overview} for clarity).
By replacing $\mathbf{x}_{u,i,\mathbf{z}}$ in equation \eqref{eq:fm}, and expanding $w_{j,j'}$, the individual utility function $f_{\mathbf{z}}(u,i|\Theta)$ can be written as:

\begin{equation}
\label{eq:utility-f}
\begin{split}
f_{\mathbf{z}}(u,i|\Theta) = f(\mathbf{x}_{u,i,\mathbf{z}}|\Theta) = w_0 + w_u + w_i + \sum_{z \in \mathbf{z}}w_z x_z + \sum_{f=1}^{k}v_{u,f}v_{i,f} \\+ \sum_{z \in \mathbf{z}}x_z \sum_{f=1}^{k}v_{u,f}v_{z,f} + \sum_{z \in \mathbf{z}}x_z \sum_{f=1}^{k}v_{i,f}v_{z,f} 
\end{split}
\end{equation}

By replacing \eqref{eq:utility-f} in \eqref{eq:utility-g}, the pairwise utility function $g$ can be written as:
\begin{equation}
\label{eq:utility-g2}
\begin{split}
g_{\mathbf{z}}(u,i,j|\Theta) = w_i - w_j + \sum_{f=1}^{k}v_{u,f}(v_{i,f}-v_{j,f}) + \sum_{z \in \mathbf{z}}x_z \sum_{f=1}^{k}v_{z,f}(v_{i,f}-v_{j,f}) 
\end{split}
\end{equation}

Now we define the FM-Pair objective function by taking the logarithm of the likelihood function in equation \eqref{eq:bpr-ll} and adding regularization terms:
\begin{equation}
\label{eq:objective2}
L(\Theta,S_P)=\sum_{(u,i,j) \in S_P} \ln \sigma(g_{\mathbf{z}}(u,i,j|\Theta)) - \sum_{\theta \in \Theta} \lambda_{\theta} \theta^2
\end{equation}

Since the FM-Pair objective function is based on likelihood, the optimal parameters are found by maximizing this function, i.e., $\Theta_{OPT}=\text{argmax}_{\Theta}L(\Theta,S_p)$.

To find the optimal parameters, optimization is done with Stochastic Gradient Descent (SGD) technique. First the parameters are initialized and then they are updated by iterating over the tuples $(u,i,j) \in S_P$ using the following update rule:
\begin{equation}
\label{eq:update}
\theta \leftarrow \theta - \eta \frac{\partial L(\Theta,S_P)}{\partial \theta}
\end{equation}
where $\eta$ is the learning rate. By replacing \eqref{eq:objective2} in \eqref{eq:update}, the update rule would be:

\begin{equation}
\theta \leftarrow \theta + \eta (\frac{e^{g_\mathbf{z}}}{1+e^{g_\mathbf{z}}} 	\frac{\partial g_\mathbf{z}}{\partial \theta} + \lambda_{\theta} \theta)
\label{eq:update-rule}
\end{equation}

Based on equation \eqref{eq:utility-g2}, the gradients of $g_\mathbf{z}$ with respect to $\theta$ is defined as:

\begin{equation}
\label{eq:gradient}
  \frac{\partial g_{\mathbf{z}}}{\partial \theta}=\begin{cases}
    1 & \text{if $\theta=w_i$}\\
    -1 & \text{if $\theta=w_j$} \\
    v_{i,f} - v_{j,f} & \text{if $\theta=v_{u,f}$}\\
    v_{u,f} + \sum_{z \in \mathbf{z}}x_z v_{z,f} & \text{if $\theta=v_{i,f}$} \\
    -v_{u,f} - \sum_{z \in \mathbf{z}}x_z v_{z,f} & \text{if $\theta=v_{j,f}$} \\ x_z (v_{i,f}-v_{j,f}) & \text{if $\theta=v_{z,f}$} \\
    0 & \text{otherwise}
  \end{cases}
\end{equation}

The parameters $w_i$ are typically initialized by 0 and the factorization parameters $v_{*,f}$ should be initialized by a zero-mean normal distribution with standard deviation $\sigma_0$ for a better performance. The parameter $\sigma_0$ is one of the hyper-parameters of SGD that typically is tuned by cross-validation or by using a validation set.

The SGD algorithm typically iterates over the entire training data and updates the parameters according to the update rule. \cite{Rendle09} suggests to draw the positive feedback from the input data by bootstrapping with replacement to prevent consecutive updates on a same user or item for faster convergence. FM-Pair first draws a positive feedback $(u,i,\mathbf{z}_{u,i})$ from the input dataset $D$ and then samples a negative item $j$ from $I\backslash I^+$ uniformly. In the next step, the utility function $g$ is calculated and then the parameters are updated according to the update rule \eqref{eq:update-rule}. Figure \ref{alg:pair-fm} summarizes the FM-Pair learning algorithm. 

\begin{figure}
	\begin{algorithmic}[1]
		\Procedure{Learn FM-Pair}{$D$}
		\State initialize $\Theta$ 
		\Repeat 
		\State sample $(u,i)$ from $D$ and create $\mathbf{x}_{u,i,\mathbf{z}}$
		\State sample $j$ from $I \backslash I^+_u$ create $\mathbf{x}_{u,j,\mathbf{z}}$
		\State let $g_{\mathbf{z}}(u,i,j|\Theta) = f(\mathbf{x}_{u,i,\mathbf{z}}|\Theta) - f(\mathbf{x}_{u,j,\mathbf{z}}|\Theta)$
		\State update $\Theta$ according update rule \eqref{eq:update-rule}
		\Until{convergence}
		\State \Return $\Theta$
		\EndProcedure
	\end{algorithmic}
	\caption{Learning FM-Pair with Stochastic Gradient Descent.}
	\label{alg:pair-fm}
\end{figure}

\subsection{Computational Complexity}
\label{sec:complexity}
FM-Pair have a linear learning and prediction time complexity. The main effort in the FM-Pair SGD algorithm is to calculate $g_{\mathbf{z}}(u,i,j)$ (line 6 in Figure \ref{alg:pair-fm}). According to \eqref{eq:utility-g2}, this can be done in $\mathbb{O}(k+|\mathbf{z}|k)$. Sampling positive pairs $(u,i)$ and negative items $j$ (lines 4 and 5 in Figure \ref{alg:pair-fm}) can be done efficiently~\cite{Rendle14} in $\mathbb{O}(1)$. Updating the parameters are done according to \eqref{eq:update-rule} and \eqref{eq:gradient}. For each point $(u,i,j,\mathbf{z})$ in the training data only the parameters corresponding to that point is updated since the gradient of the other parameters are 0. Thus the complexity of updating the parameters for each training point (line 7 in Figure \ref{alg:pair-fm}) is $\mathbb{O}(|\mathbf{z}|k)$ according to \eqref{eq:gradient}.
Putting it all together, the complexity of one iteration in FM-Pair SGD algorithm is $\mathbb{O}(k(|\mathbf{z}|+1))$ and the complexity of a full iteration on the entire training data $D$ is $\mathbb{O}(k(|\mathbf{z}|+1)|D|)$. Therefore, the computational complexity of FM-Pair is linear in terms of number of latent factors $k$ and number of auxiliary features $\mathbf{z}$. In the experiments section we empirically demonstrate the training time of FM-Pair based on $k$ and $\mathbf{z}$.

\subsection{Analogy Between FM-Pair and BPR-MF}
\label{sec:analogy}
FM-Pair can mimic other factorization methods with feature engineering, similar to the standard Factforization Machines. A specific model that can be represented with FM-Pair is BPR-MF (BPR with Matrix Factorization utility)~\cite{Rendle09}.  
The matrix factorization model calculates the utility of a user-item interaction as the inner product of the user and item factors, that it, $f^{MF}(u,i) = \sum_{f=1}^{k} v_{u,f}v_{i,f}$. By considering $f^{MF}$ as the utility function of matrix factorization model, the utility of triples $(u,i,j)$ in BPR-MF is defined as:
\begin{equation}
\label{eq:utility-mf}
g^{MF}(u,i,j) = \sum_{f=1}^{k}v_{u,f}(v_{i,f}-v_{j,f})
\end{equation}

By comparing the above equation with \eqref{eq:utility-g2}, one can notice that $g^{MF}$ is a special case of $g_{\mathbf{z}}(u,i,j)$ where $\mathbf{z} = \emptyset$ (i.e., no auxiliary features) and the parameters $w_i$ and $w_j$ are $0$. In fact when there are no auxiliary features, FM-Pair, compared to BPR-MF, learns two additional parameters $w_i$ and $w_j$ that can be regarded as global item biases for positive and negative items.

%- Discuss about complexity of the model?

\section{Improved Recommendations with Auxiliary Data}
\label{sec:aux}
A great advantage of Factorization Machines as mentioned earlier, is that they are capable of exploiting arbitrary information as auxiliary features to improve recommendations. In this section, we propose to apply FM-Pair for the context-aware and cross-domain collaborative filtering for datasets with implicit feedback.
%\citet{Rendle11_2} propose context-aware recommendation with FMs and \cite{Loni2014} propose cross-domain recommendation with FMs. Both studies exploit auxiliary features as
%However, both studies are limited only to data with explicit feedback. With the adaptation that we proposed in the previous section, FMs can also be used for implicit data with unary feedback and the auxiliary information can also be used to improve recommendations for this type of data. In this section we discuss in more details how FMs can be improved by auxiliary data (not only context) for both explicit and implicit data and we further introduce some feature engineering method that can exploit factorization machines more effectively. 

\subsection{Context-Aware Recommendation with FM-Pair}
Context is a dynamic set of parameters describing the state of the user at the moment of experience. Some studies also consider user and item attributes as a type of context~\cite{adomavicius11}. Context-aware recommendation relies on these additional information to better learn from the interactions between users and items. In FM-Pair, context can be considered as one type of auxiliary features for user-item interactions. We represent the context of a user feedback with the following sparse representation:
\begin{equation}
\label{eq:z}
\mathbf{z}(u,i)=\{(z,x_z) | x_z \neq 0 \} 
\end{equation}
where $z$ is a context and $x_z$ is its corresponding value. For example if available context of an interaction $(u,i)$ are user mood (e.g. ``happy") and movie genre (e.g. ``action"), then we can represent the context of interaction with $\mathbf{z}(u,i)=\{ (\text{happy},1), (\text{action}, 1) \}$.
By expanding the feature vector $\mathbf{x}$ with context features, the feature vector $\mathbf{x}$ would have the following sparse form:
\begin{equation}
\label{eq:x-sparse}
\mathbf{x}(u,i,\mathbf{z}) = \mathbf{x}_{u,i,\mathbf{z}} = \{(u,x_u),(i,x_i)\} \cup \mathbf{z}(u,i)
\end{equation}
and the following expanded form:
\begin{equation}
\label{eq:x-extended}
\mathbf{x}_{u,i,\mathbf{z}} = (\underbrace{0,\dots,0,x_u,0,\dots,0}_{\mbox{\scriptsize{$|U|$}}},\underbrace{0,\dots,0,x_i,0,\dots,0}_{\mbox{\scriptsize{$|I|$}}}, \underbrace{x_{z_1},\dots,x_{z_m}}_{\mbox{\scriptsize{$|Z|$}}})
\end{equation}
where $Z$ is the set of contextual feature and $m=|Z|$.
The parameters $x_u$, $x_i$ and $x_z$ are the actual values of features, which be seen as weight parameters specifying the strength of features. The parameters $x_u$ and $x_i$ are typically considered to be 1 to just indicate the presence of features. The parameters $x_z$ can be assigned with any real-values to control the weight of contextual features. For the categorical context such as user mood typically binary values are used similar to $x_u$ and $x_i$. However, the binary values can also be replaced with real values. 
%In that case those values can be seen as a weight parameter that influences on the utility of $\mathbf{x}_{u,i,\mathbf{z}}$ (see equation \eqref{eq:fm}).
If the features are continues by their nature (e.g. user age, time of the day and number of clicks), a real value can be considered as the value of the feature.
According to~\cite{Rendle11_2} it is often desirable that the value of auxiliary features sum up to 1. Continuous features can also be mapped to categorical features by forming several bins. With our preliminary experiments we found that mapping continuous context to categorical features results to better accuracy. In Section \ref{sec:experiments}, we describe our feature mapping on the two datasets with contextual features.

\subsection{Cross-Domain Recommendations}
\label{sec:cd}
Cross-Domain Collaborative Filtering (CDCF) methods exploit additional information from \textit{source}\footnote{The source domains are also referred as ``auxiliary" domains. In this work we used the term ``source" to avoid confusion with auxiliary features.} domains to improve recommendations in a \textit{target} domain. The main idea of CDCF is that a user's taste in one domain (e.g., movie) can be exploited to better learn user's taste on another domain (e.g., music). Different methods for the problem of CDCF have been proposed. A great overview of the existing solutions can be found in~\cite{Cantador15}. Among different techniques that have been developed for cross-domain CF, in an earlier study~\cite{Loni2014} we proposed a method for the problem of CDCF with Factorization Machines for the rating prediction problem.
In this section, we propose a similar technique that has been adapted for FM-Pair and thus can be applied for datasets with implicit feedback.

Thanks to the flexibility of Factorization Machines in exploiting additional features, the information from source domains can be translated to auxiliary features to expand the feature vectors in the target domain. The expanded feature vectors can be then used as the input data for FM-Pair to train a model. With the same assumption as the case of context-aware recommendation, the extra features can enrich the feature vectors to better learn the user preferences. We refer to our proposed cross-domain CF method with FM-Pair as FM-Pair-CD. The advantage of FM-Pair-CD is that it does not require any adaptation to the underlying FM-Pair model and thus the model can be used out-of-the-box. FM-Pair-CD proposes a simple and yet effective way to exploit features from source domains in order to improve the performance of recommendations in the target domain. 
Here the domain refers to the type of item that user interacted with. For example if user provides a feedback for a movie, the domain is ``movies". 

To understand how FM-Pair-CD represents the auxiliary features, suppose $p$ is the number of source domains and $I_j(u)$ is the set of items in domain $j$ that user $u$ interacted with. FM-Pair-CD proposes to use one auxiliary feature for every item that user interacted with in the source domains. Therefore, the feature vectors $\mathbf{x}$ in the target domain can be represented with the following sparse form consisting of both target and source domain features:
\begin{equation}
\label{eq:x-slice}
\mathbf{x}(u,i) = \{\underbrace{(u,1),(i,1)}_{\mbox{\scriptsize{target domain features}}},\underbrace{\cup_{j=1}^{p} \{ (z,x_z(u,j)) | z \in I_j(u)  \} }_{\mbox{\scriptsize{source domains' features}}}\}
\end{equation}
where $x_z(u,j)$ is the value of feature $z$, i.e., the weight that should be considered for the interaction of user $u$ with item $z$ in the source domain $j$. 
We propose the following two approaches to define the feature values $x_z(u,j)$:
\begin{itemize}
	\item Binary indicator: In this case similar to the representation of user and item, the presence of an source domain feature is specified by indicator value of 1. We denote  $x^B_z(u,j)$ as binary representation of features. In other words, in this case, with binary values we indicate which items have been rated by the user in the source domains.

	\item Normalized by count: In this case the values of source domain features are normalized by the number of feedback that user provided in the source domain. The normalized value $x^C_z(u,j)$ is defined by:
	\begin{equation}
	\label{eq:x-C}
	x^C_z(u,j)=\frac{1}{|I_j(u)|}
	\end{equation}
	
\end{itemize}

\begin{figure}[t]
	\centering
	\includegraphics[width=0.75\textwidth]{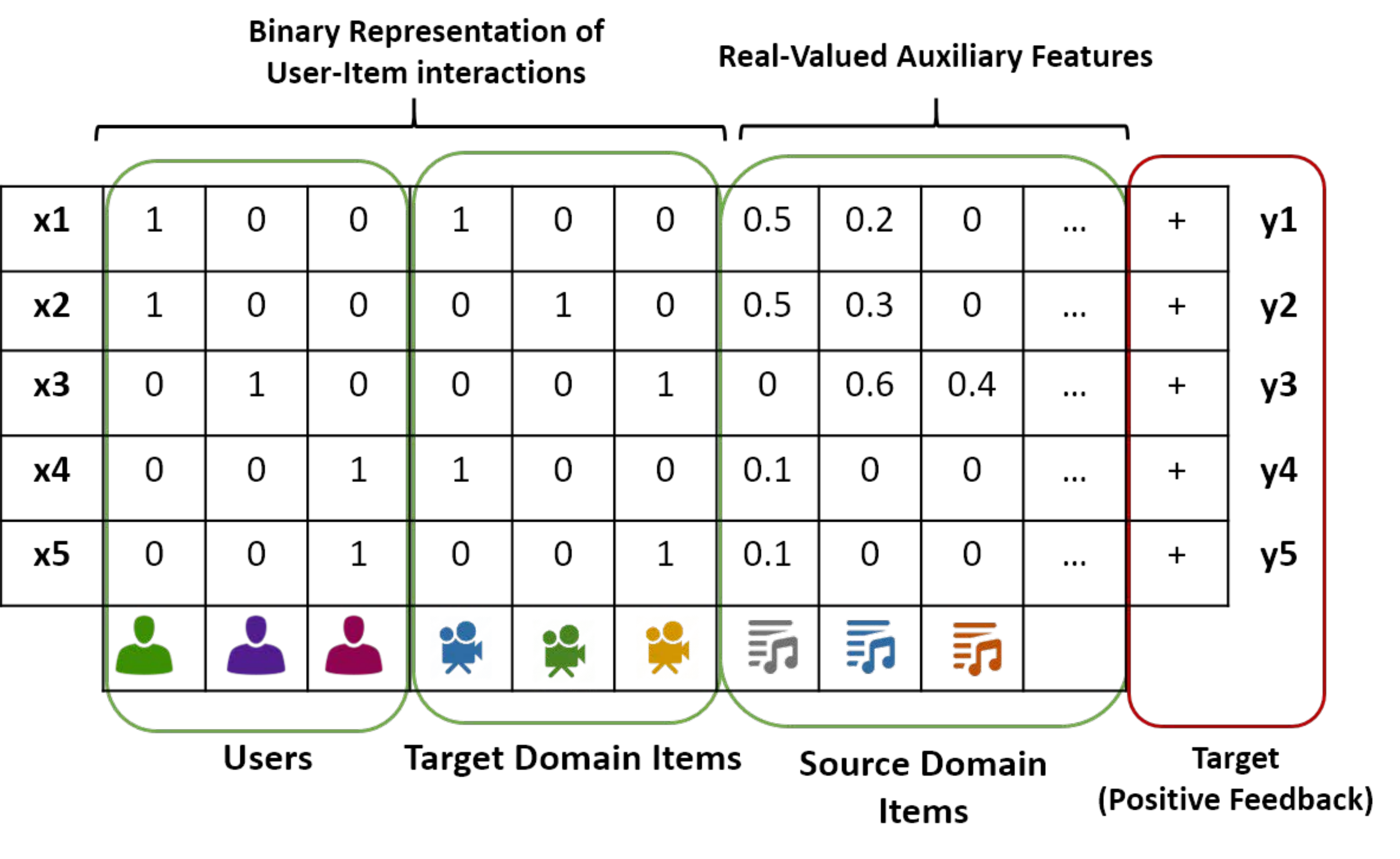}
	\caption{Representation of feature vectors in FM-Pair-CD.}
	\label{fig:cd}
\end{figure}

The auxiliary features in the FM-Pair-CD method, are in fact the items in source domains with feedback from the user. The normalization of auxiliary features ensures that the target features are not dominated by auxiliary features. Figure \ref{fig:cd} illustrates how FM-Pair-CD represents feature vectors for input data. In this example the target domain is the ``movie" domain and source domain is the ``music" domain. The music items that same user has interacted with are considered as auxiliary features for the interactions in the target domain. 
%Since for each item in the source domain one auxiliary feature is considered, the total number of features in a cross-domain problem with $p$ source domains would be:
%\begin{equation}
%	n = |U| + |I| + \sum_{j=1}^{p} |I_j|
%\end{equation}
%where $I_j$ represents the items in domain $j$. \citet{Rendle12} showed that the complexity of FMs depends on the number of non-zero elements in the input feature vectors. 
As you might notice from Figure \ref{fig:cd}, the total number of auxiliary features is the number of items in source domains. We found that using only a fraction of user feedback from source domains results in almost same improvement compared to the case that all user feedback from source domains are used. 
Such feature selection makes the size of feature vectors smaller and thus the training and prediction become faster. The selected features can be based on most popular items, highly rated items, or just randomly chosen items. In fact, by applying this feature selection method, the auxiliary features in \eqref{eq:x-slice} can be represented as:
\begin{equation}
\label{eq:aux}
\cup_{j=1}^{p} \{ (z,x_z(u,j)) | z \in I_j(u) \land z \in I^S(u) \}
\end{equation}
where $I^S(u)$ are the selected features. Since finding the most informative items from source domains is not the focus of this work, we just consider $I^S(u)$ to be random items from source domains.
%We will show the effect of such feature selection method in performance and complexity of the model in the experiments section.

\section{Datasets, Experiments and Evaluation}
\label{sec:experiments}
In this section we first introduce the datasets that we used in this work, then we describe our evaluation method and then we explain our experiments.

\subsection{Datasets}
We used the following four datasets in this work, ranging from the popular MovieLens dataset to a recent industry dataset of XING. These datasets are chosen so that we can test different scenarios with auxiliary features and to cover both implicit and explicit feedback scenarios.
\begin{itemize}
\item \textbf{MoviLens 100K}: The MovieLens 100K dataset\footnote{http://grouplens.org/datasets/movielens/} is a popular benchmark dataset for recommender systems with explicit feedback consisting of ~100K user rating on the scale of 1 to 5. This dataset also contains several user and item attributes that can be used as auxiliary features in FM-Pair.

\item \textbf{Amazon}: The Amazon dataset \cite{Leskovec07} consists of user ratings on the products on the Amazon website. The ratings are on the same scale as the MovieLens dataset. The products belong to four different domains: Books, Music CDs, Video tapes and DVDs. This dataset has been used for some previous work on cross-domain CF~\cite{Loni2014,Hu13}.

\item \textbf{Frappe}: Frappe is a context-aware mobile app discovery tool. It logs number of time users run an application on their mobile phone. It also logs several contexts such as time, date, location and weather. The Frappe dataset \cite{frappe15} consists of ~100K implicit positive feedback. An observation is considered as a positive feedback if user runs an application at least one time. We used this dataset because of the presence of several contextual features. This dataset has also been used in one of the related work~\cite{Nguyen14}.

\item \textbf{XING}: XING\footnote{http://www.xing.com/} is a professional social network and a job discovery platform. This dataset is a subset of the RecySys 2016 challenge\footnote{http://2016.recsyschallenge.com/} and consists implicit user feedback on job postings. Users can either click on a job posting, bookmark it or apply for it. We consider any user action on a job positing as a positive feedback. Since the original dataset was extremely sparse, we densified the dataset by considering users and items with at least 20 interactions.
\end{itemize}

Table \ref{tab:datasets} list the statistics of the three datasets that we used in this work.

\begin{table}
	\centering
	\caption{Statistics of the dataset used in this work}
    \begin{tabular}{p{3cm}lllll}
    \hline \hline
    \textbf{Dataset} & \textbf{\#Users} & \textbf{\#Items} & \textbf{\#Feedback} & \textbf{Sparsity(\%)} & \textbf{Scale} \\ \hline
    MovieLens 100K & 943 & 1,682 & 100K & 93.74 & 1-5\\ 
    Amazon & 15,994 & 84,508 & 270K & 99.98 & 1-5 \\ 
    Frappe & 957 & 4,082 & 96K & 97.54 & Implicit \\ 
    XING & 9,751 & 9,821 & 223K & 99.76 & Implicit \\ \hline \hline
    \end{tabular}
    \label{tab:datasets}
\end{table}

\subsection{Experiments Setup and Evaluation}
\label{sec:eval}
All experiments in this work are done with four-fold cross-validation to make sure the hyper-parameters are not tunned for one particular test set.  
FM-Pair is implemented as a part of the WrapRec\cite{Loni2014_2} open source project. WrapRec can be used as a command line tool in different platforms. The source code and documentation can be found in \url{http://wraprec.crowdrec.eu/}.

%We used different evaluation metrics for explicit and implicit feedback scenarios. For rating prediction problems, we used RMSE and MAE evaluation metrics. For unary feedback scenarios, ranked lists are generated by scoring items and then three rank-based metrics are used: Recall, MRR (Mean Reciprocal Rank) and NDCG (Normalized Discounted Cumulative Gain).

\subsubsection{Evaluation Method}: The performance of our experiments are evaluated with two ranking metrics namely Recall and NDCG (Normalized Discounted Cumulative Gain). To calculate these metrics on datasets with positive-only feedback, typically for each user in the test set a rank list is created. The metrics are then calculated based on the presence of test feedback on top of the list. However, when auxiliary features such as context are available for the feedback, this strategy is not suitable as it is not clear based on which context the scores (i.e., the utility of a user-item interaction) should be generated. To make an unbiased estimation of performance when auxiliary features are available, we applied the approach known as \emph{One-plus-random} \cite{Cremonesi10,Bellogin11} where the following procedure is applied:
For every user-item interaction in the test set (which is a positive feedback with possible auxiliary features), 1000 random items which are not observed with that user are scored based on the same user and auxiliary features. Then a ranked list including the target item is created. The position of the target item in the ranked list is used to calculate the metrics.
If the targeted test point appears in the top N positions of the list, then it would be considered as a \emph{hit}. In case of a hit, the recall of a single test point would be 1 and otherwise it would be 0. 
%Similarly, precision of a single point would be $1/N$ in case of a hit and 0 in case of a miss. 
The overall metric is calculated by averaging on all points. That is:
\begin{equation}
\text{Recall@N} = \frac{1}{|T|} \sum_{i=1}^{|T|} \mathbb{I}(r_i \leq N) %=\frac{\text{\#hits}}{|T|}
\end{equation}
where $r_i$ is the rank of the $i$th test sample in the generated list, $\mathbb{I}$ is the indicator function and $|T|$ is the size of test set. The above metric can be interpreted as the hit rate of the test samples where a hit is defined as presence of the relevant test point in the top N positions of the ranked list. %Similarly, precision can be calculated:
%\begin{equation}
%\label{eq:prec}
%\text{Precision@N} = \frac{1}{|T|} \sum_{i=1}^{|T|} \frac{1}{N} \mathbb{I}(r_i %\leq N)=\frac{\text{Recall@N}}{N}
%\end{equation}

Based on the one-plus-random evaluation method, we also adopted the MRR metric as follows:
\begin{equation}
\text{MRR@N}=\frac{1}{|T|} \sum_{i=1}^{|T|} \frac{1}{r_i} \mathbb{I}(r_i \leq N)
\end{equation}
%\begin{equation}
%\text{DCG@N}=\frac{1}{|T|} \sum_{i=1}^{|T|} \frac{1}{\log_2(r_i+1)} \mathbb{I}(r_i \leq N)
%\end{equation}

%The NDCG metric is calculated by normalizing DCG. Normalization is done by divining DCG@N by the Idealized DCG (IDCG@N)\footnote{https://www.kaggle.com/wiki/NormalizedDiscountedCumulativeGain}. %Since Precision@N can be calculated based on Recall@N (see \eqref{eq:prec}) we only report Recall@N in our results. 
We use the MRR metric since it also takes into account the position of the relevant item in the list.
Note that these metrics are not absolute metrics \cite{Bellogin11} and their value does not reflect the real performance of the system. However, they are reliable metric to \emph{compare} the performance of different experiments.

\subsection{Comparison of FM-Pair with Other Methods}
The proposed FM-Pair algorithm is compared with several methods on the four datasets. In this experiment no auxiliary or context features are used and the methods are compared only based on the positive feedback in the user-item matrix. The following setups have been tested in this experiment:

\begin{itemize}

\item \textbf{Most-Popular}: This is a baseline method where the most popular items are recommended to the users. 

\item \textbf{FM-Map}: In this setup the training is done similar to the FMs for  rating prediction. For the positive feedback in the training set the output value of +1 is considered. Same number of unobserved interactions are sampled uniformly and they are considered as negative feedback with output value of -1. The positive and sampled negative feedback is used to train the FMs model.
To resemble the experiment for datasets with explicit feedback, the ratings higher than user's average were mapped to +1 and negative feedback were sampled in the same way as the implicit feedback datasets.
%\item \textbf{FM-RatingBased}: This setup is only applicable for dataset with explicit feedback. Here the standard FMs model is used for predicting the ratings from which a ranked list is created.

\item \textbf{BPR-MF}: This method is an implementation of BPR method with Matrix Factorization as the utility function~\cite{Rendle09}. With this method there is no possibility to incorporate auxiliary information.

\item \textbf{FM-Pair}: This is the proposed method in this work. The FM-Pair algorithm is listed in Figure \ref{alg:pair-fm}. For the two datasets of MovieLens and Amazon with explicit feedback, the ratings above user's average rating is considered as positive feedback.

%\item \textbf{GPFM}: This setup refers to Gaussian Process Factorization Machines (GPFM)~\cite{Nguyen14} with the pairwise optimization model. 

\end{itemize}

\subsubsection{Hyper-parameters and Experimental Reproducibility}
\label{sec:rep}
The three datasets of MovieLens, Amazon and Frappe are publicly available. For the experiments in this section, the number of factors (parameter $k$) is set to 10 and the number of iteration of the SDG algorithm on the training data is set to 300.
The standard deviation of the the normal distribution for initialization (parameter $\sigma_0$) is set to 0.1. The learning rate of the SGD algorithm (parameter $\lambda$) varies per dataset. The following learning rates are used for each dataset: XING: 0.075, Frappe and MovieLens: 0.005, Amazon: 0.001.
The two hyper-parameters of $\sigma_0$ and $\lambda$ are found with a grid search with our four-fold cross-validation experiments. That is, the hyper-parameters that result to the best performance on the \textit{average} of all folds are chosen, thus they are not optimized for one particular test set.

\subsubsection{Description of the Results} Table \ref{tab:fm-pair} lists the performance of the above five methods on the four datasets based on Recall@10 metric\footnote{Other ranking metrics such as MRR and NDCG are also calculated, but due to the high correlation between the values, we only report Recall@10 for this experiment.}. As it can be seen from the table, in three out of the four datasets, the FM-Pair is performing better than the other baselines. 
In the XING dataset, BPR-FM is slightly performing better than FM-Pair. It is worth mentioning  that when there are no auxiliary features (such as this experiment) the underlying model of FM-Pair is very similar to BPR-MF (see Section \ref{sec:analogy}). This can explain the close performance of the two methods. Nevertheless, the additional parameters of FM-Pair contribute to some accuracy gain in three out of the four datasets. 
Another observation that you can see in this table, is that the FM-Map method is not really effective for ranking compared to the two pairwise methods in our experiments. This can be explained by the fact that the standard FMs optimization method is a pointwise method that in principle learns to correctly predict the mapped scores and it is not optimized for ranking. Note that we also tried to map the sampled unobserved feedback to other values than -1, but in practice the result were very similar.
For the two datasets of Amazon and MovieLens with explicit feedback, we did an additional experiment where the model is trained with the standard FM with original ratings. The ranked lists are then generated based on the predicted ratings. For this experiment we achieved a Recall of 0 for Amazon and 0.001 for the MovieLens dataset. This shows that even for datasets with explicit feedback, ranking based on predicting ratings is not really effective. Previous studies~\cite{Balakrishnan12,Cremonesi10} also showed the ineffectiveness of pointwise methods for ranking.
%The results of the other two methods that are based on the standard Factorization Machines (FM-RatingBased and FM-Map) clearly show that standard FMs are not optimized for ranking tasks. Even for datasets with explicit ratings, ranking based on the predicted ratings is not an effective approach for generating a ranked list. 

\begin{table}[t]
	\centering
	\caption{Comparison of different learning-to-rank methods on the four dataset based on Recall@10. The numbers in parentheses are standard deviations of the results in the four folds.}
	\begin{tabularx}{\textwidth}{@{\extracolsep{\fill}}lllll}
		\hline \hline
		\textbf{Method / Dataset} & \textbf{XING} & \textbf{Frappe} & \textbf{Amazon} & \textbf{MovieLens} \\ \hline
		Most-Popular & 0.0306 \scriptsize{(0.0005)} & 0.1643 \scriptsize{(0.0087)} & 0.0222 \scriptsize{(0.0010)} & 0.1180 \scriptsize{(0.0012)} \\ 
		%FM-RatingBased & NA & NA & 0.0000 \scriptsize{(0.0000)} & 0.0010 \scriptsize{(0.0000)} \\
		FM-Map & 0.0287 \scriptsize{(0.0015)} & 0.1230 \scriptsize{(0.0260)} & 	0.0348 \scriptsize{(0.0014)} & 0.0728 \scriptsize{(0.0036)} \\
		BPR-MF & 0.2925 \scriptsize{(0.0030)} & 0.1428 \scriptsize{(0.0167)} & 0.0962 \scriptsize{(0.0056)} & 0.2278 \scriptsize{(0.0024)} \\ 
		FM-Pair (this work) & 0.2920 \scriptsize{(0.0077)} & 0.1816 \scriptsize{(0.0161)} & 0.0972 \scriptsize{(0.0030)}  & 0.2357 \scriptsize{(0.0016)} \\ \hline
	\end{tabularx}
	\label{tab:fm-pair}
\end{table}

\subsubsection{Comparison with GPFM}
\begin{figure}[t]
	\centering
	\includegraphics[width=0.75\textwidth]{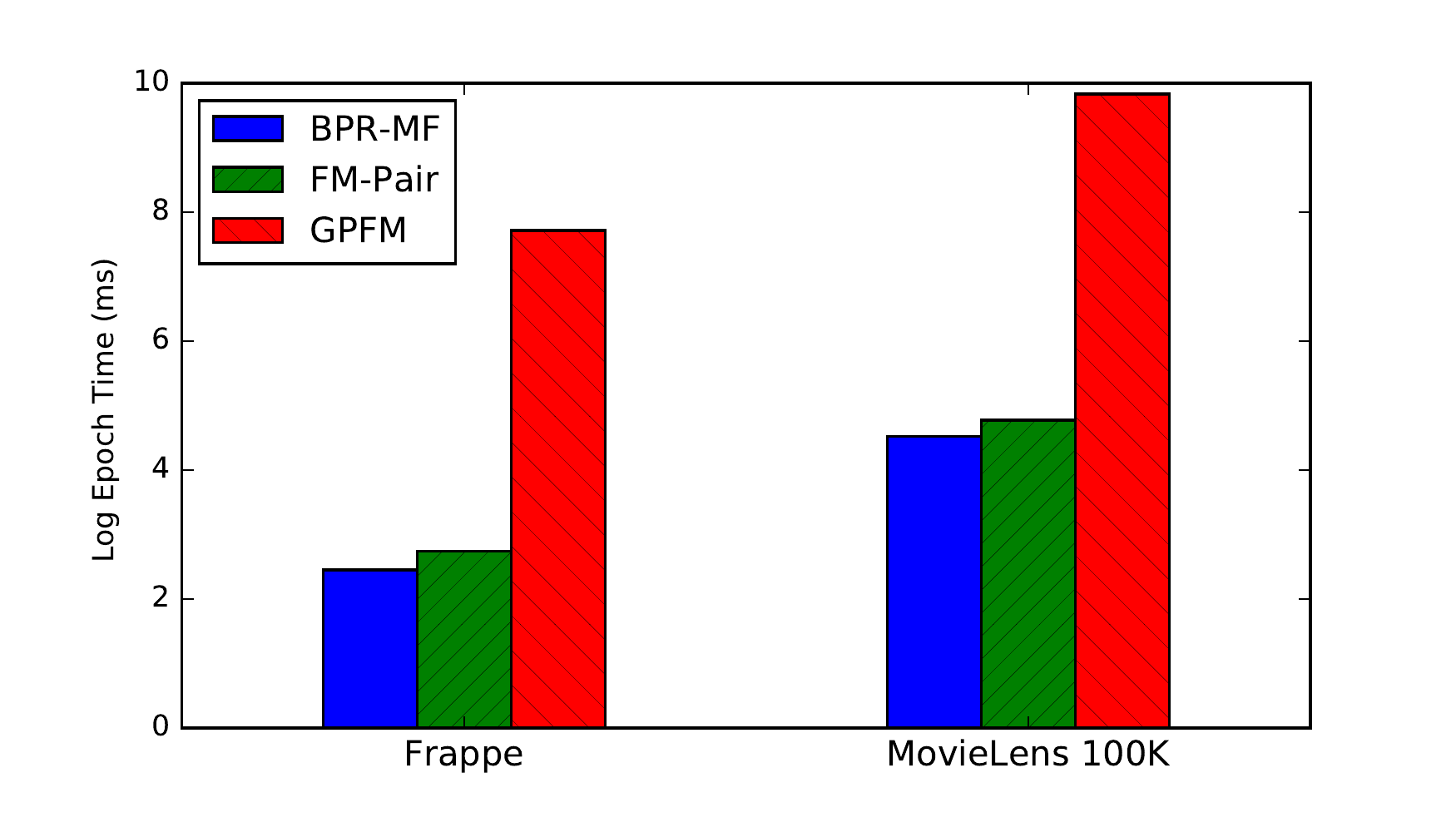}
	\caption{Comparison of the epoch time (the time of a full iteration on the dataset in milliseconds) of three pairwise learning-to-rank methods on the log scale.}
	\label{fig:epochtime}
\end{figure}

: In addition to the methods listed in Table \ref{tab:fm-pair}, we also compare the performance of FM-Pair with the pairwise method of GPFM~\cite{Nguyen14} since it is very close to our work as it also adapted a pairwise optimization technique for FMs. 
We tested the GPFM method on Frappe and MovieLens datasets. For the Frappe dataset we achieved a Recall@10 of 0.1615 and for the MovieLens a Recall@10 of 0.1562  was achieved, both less than the performance of FM-Pair (See Table \ref{tab:fm-pair}).
However, the remarkable advantage of FM-Pair compared to GPFM is that the computational complexity of FM-Pair is significantly lower than GPFM. Figure \ref{fig:epochtime} compares the epoch time (the time of a full iteration on dataset) of the three methods of BPR-MF, GPFM and FM-Pair on two datasets of Frappe and MovieLens. The numbers are represented in the log scale due to the significant difference between the epoch time of GPFM with the other two methods. The epoch time of FM-Pair is slightly higher than the epoch time of BPR-MF due to the presence of two additional parameters (see Section \ref{sec:analogy}). The GPFM method on the other hand, is significantly slower that the other two methods due to the fact that GPFM need to calculate the inverse of the covariance matrix of the preference kernels~\cite{Nguyen14}. This introduces a significant computational complexity in the training of GPFM.

Due to the high space complexity of GPFM, running the experiment on the larger datasets of XING and Amazon was not even possible on our testing machine\footnote{The experiments are run on a machine with 8 GB of memory and an Intel i5 processor with 4 CPU cores.}. Since the complexity of GPFM is significantly higher than the other methods, we did not further investigate on testing GPFM on our larger datasets.

We used the Matlab implementation\footnote{\url{http://trungngv.github.io/gpfm/}} of GPFM that was released with that work to train the model but the evaluation was done in the same way as other methods, as described in Section \ref{sec:eval}, to have a fair comparison between the methods. The kernel of the Gaussian process is chosen to be the RBF kernel, the recommended kernel of the model.

%an additional factor of $a^2$ in the complexity of this method, where $a$ is the average number of user feedback. This means that if on a dataset on average users have 10 feedback, the training time is 100 time slower than the other two methods.

\subsection{FM-Pair with Auxiliary Data}
In the second set of our experiments, we test the performance of FM-Pair with auxiliary information. We use FM-Pair for the two scenarios that we described in Section \ref{sec:aux}: context-aware recommendation and cross-domain collaborative filtering.

\subsubsection{FM-Pair with Context}
Among the four datasets that we use in this work the two datasets of Frappe and MovieLens have several context and attributes. The final context and attributes that are used in the experiments are found with a naive greedy method, where the top performing features are combined. 
For the Frappe dataset the following contexts are used: daytime (with seven possibilities of \texttt{sunrise, morning, noon, afternoon, sunset, evening, night}), weekday (day of the week), isweekend (with two values of \texttt{workday} or \texttt{weekend}) and homework (with three values of \texttt{unknown, home, work}). For the MovieLens dataset, we used the genre of movies as auxiliary features in FM-Pair. Each movie in this dataset has one or more genres from the 17 genres that exists in this dataset.

Table \ref{tab:context} compares the performance of FM-Pair with context or attributes as auxiliary features (FM-Pair-Context), with the original FM-Pair without any auxiliary features. We reported Recall@10 and MRR@10 for the two setups in this experiment. The results show that the FM-Pair can clearly exploit context or auxiliary attributes if they are present in a dataset.

\subsubsection{Cross-Domain Recommendation with FM-Pair}
\begin{figure}[b]
	\centering
	\includegraphics[width=\textwidth]{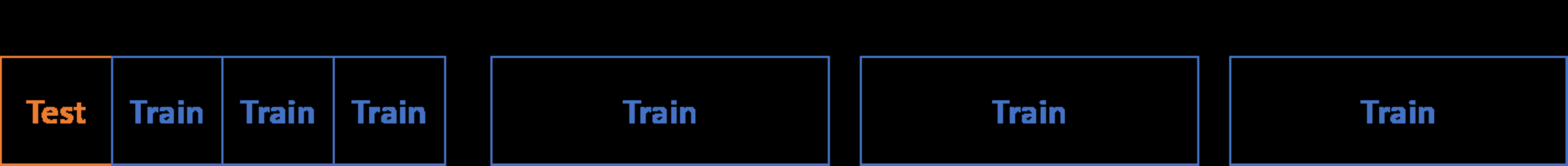}
	\caption{Illustration of the cross-validation splitting method for the cross-domain recommendation experiment with FM-Pair on the Amazon dataset.}
	\label{fig:cd-cv}
\end{figure}

To test the performance of FM-Pair for cross-domain collaborative filtering, we use the dataset of Amazon where the items come from four different domains. We use the two domain of \texttt{books} and \texttt{music} as target domains and use other domains as source domains. The experiments are done with four-fold cross-validation and on each fold only the interactions from the target domain are evaluated. The source domains are used to generate auxiliary features, as described in Section \ref{sec:cd}. Figure \ref{fig:cd-cv} illustrates our four-fold cross-validation splitting method on the Amazon dataset with one target domain and three source domains. The design choices and hyper-parameters of the experiment are the same as the ones described in Section \ref{sec:rep} except that for the \texttt{books} domain, the learning rate of the SGD algorithm is set to 0.001 due to its faster convergence.

The following three setups are used to demonstrate the performance of the FM-Pair method for cross-domain recommendations. In all setups, the evaluation is done for the target domain.

\begin{itemize}
	\item \textbf{FM-Pair}: In this setup FM-Pair is only applied on the target domain and source domains are not exploited.
	
	\item \textbf{FM-Pair-All}: In this setup the source domains are used as additional training samples. Thus, no auxiliary features are generated. 
	
	\item \textbf{FM-Pair-CD}: In this setup source domains are exploited to generate auxiliary features for the training feature vectors in the target domain. In fact, the number of training samples in this setup is the same as the first setup but the feature vectors are expanded with auxiliary features. The value of auxiliary features are defined based on equation \eqref{eq:x-C} and for each user we take at most five feedback from each source domain to avoid large feature vectors.
\end{itemize}

Table \ref{tab:cd} lists the performance of the above three setups on the Amazon dataset where domains of \texttt{books} and \texttt{musics} are used as target domains. First of all, as you can see in this table, the FM-Pair-CD outperform the other two methods on the accuracy of recommendations. The second interesting observation is that when recommendations are generated for a target domain, using only interactions from that particular domain (setup FM-Pair) is better than using the entire dataset for training (setup FM-Pair-All). Similar effect has been shown in a previous study as well~\cite{loni2015}: using a sensible subset of data can perform better than using the entire dataset. 
In fact exploiting the source domains by means of auxiliary features in FM-Pair has a better effect than blindly combining all samples from all domains in one big dataset.

\begin{table}[t]
	\centering
	\caption{Performance of FM-Pair with context compared to the standard FM-Pair without auxiliary features.}
	\label{tab:context}
	\begin{tabularx}{\textwidth}{@{\extracolsep{\fill}}l|ll|ll}
		\hline \hline
		& \multicolumn{2}{l|}{\textbf{Recall@10}} & \multicolumn{2}{l}{\textbf{MRR@10}} \\
		\hline
		\textbf{Method / dataset} & Frappe       & MovieLens      & Frappe      & MovieLens     \\ \hline
		FM-Pair &   0.1816    & 0.2357 &   0.0745  &  0.1027 \\
		FM-Pair-Context &  \textbf{0.2064}  & \textbf{0.2601} & \textbf{0.0890} & \textbf{0.1191} \\ \hline             
	\end{tabularx}
\end{table}

\begin{table}[t]
	\centering
	\caption{Performance of cross-domain recommendation with the FM-Pair-CD method compared to the single-domain training scenarios.}
	\label{tab:cd}
	\begin{tabularx}{\textwidth}{@{\extracolsep{\fill}}l|ll|ll}
		\hline \hline
		& \multicolumn{2}{l|}{\textbf{Recall@10}} & \multicolumn{2}{l}{\textbf{MRR@10}} \\
		\hline
		\textbf{Method / Target Domain} & Books       & Music      & Books      &  Music     \\ \hline
		FM-Pair (Target-only) &   0.1058    & 0.0966 &  0.0452   &  0.0356 \\
		FM-Pair-All &   0.0831    & 0.0855 &  0.0357  & 0.0378 \\
		FM-Pair-CD &  \textbf{0.1238}  & \textbf{0.1060} & \textbf{0.0490} & \textbf{0.0405} \\ \hline             
	\end{tabularx}
\end{table}

\subsection{Convergence and Complexity of FM-Pair}
\begin{figure}[t]
\centering
\includegraphics[width=\textwidth]{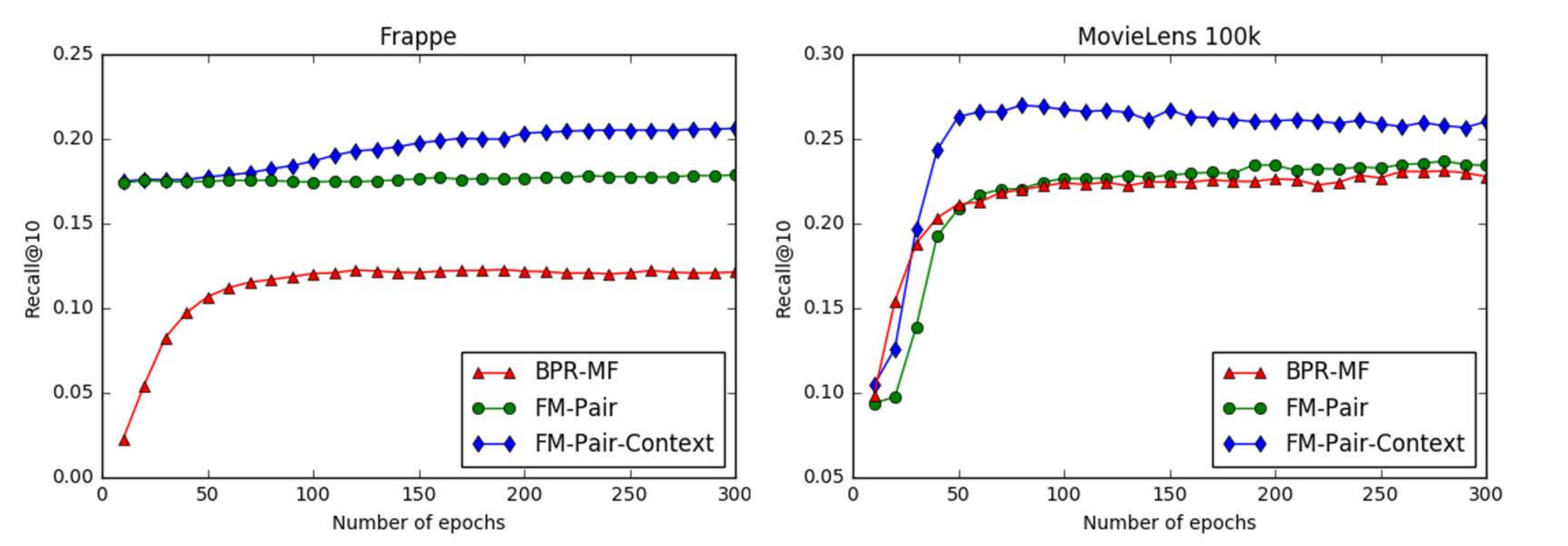}
\caption{Empirical comparison of the convergence of FM-Pair (with or without context)  with BPR-MF on the two datastes with auxiliary features.}
\label{fig:converge-context}
\end{figure}

\begin{figure}[t]
	\centering
	\includegraphics[width=\textwidth]{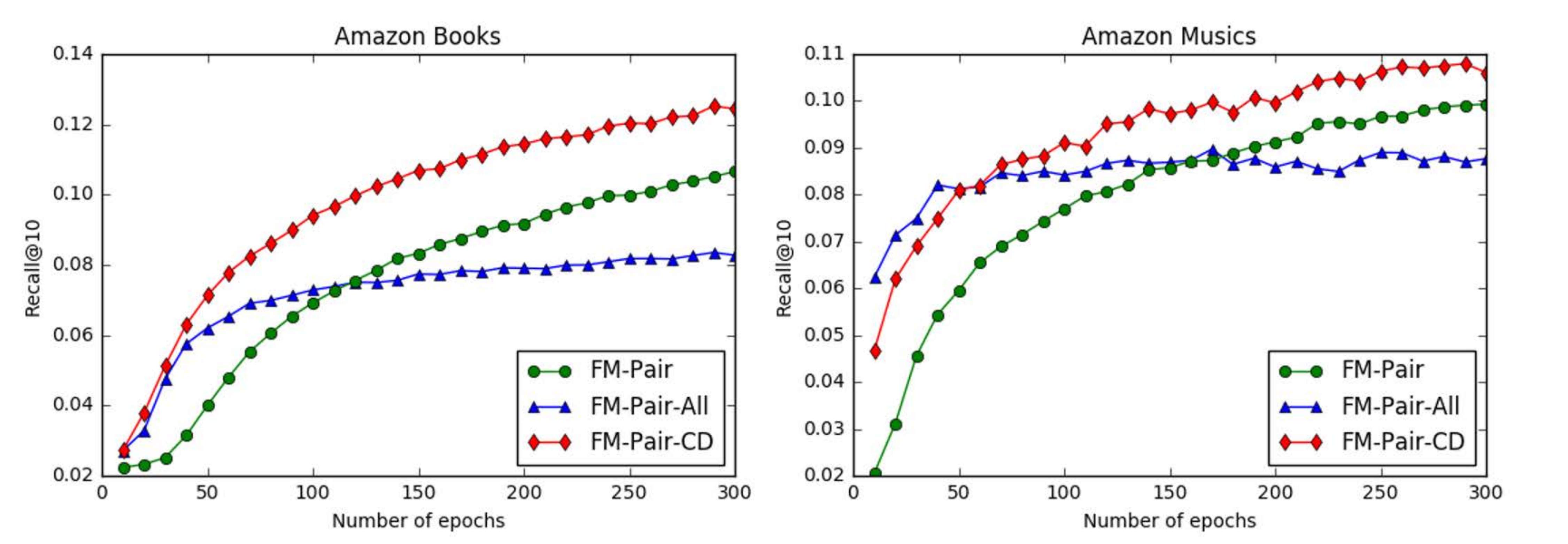}
	\caption{Empirical comparison of the convergence of cross-domain CF with FM-Pair compared to the single-domain models on two domains of the Amazon dataset.}
	\label{fig:converge-cd}
\end{figure}

In this section we further analyze the convergence and complexity of FM-Pair by monitoring the performance of FM-Pair with different number of iterations (of the training algorithm) and different dimensions of factorization. 
Figure \ref{fig:converge-context} compares the performance of FM-Pair, FM-Pair-Context and BPR-MF on different number of epochs on the two datasets of Frappe and MovieLens. In Figure \ref{fig:converge-cd}, we illustrate the performance of cross-domain recommendations with FM-Pair-CD compared to the two setups of FM-Pair and FM-Pair-All on different number of epochs. The models are evaluated on every 10 epoch with Recall@10.

As you can see in Figure \ref{fig:converge-context}, on the Frappe and MovieLens dataset all models converge rather fast due to the density of datasets. However, an interesting observation on the Frappe dataset is that FM-Pair and FM-Pair-Context already achieve a high recall after the first few epochs whereas the BPR-MF algorithm converge later and yet cannot achieve FM-Pair's performance even with higher number of epochs. A closer examination of Section \ref{sec:analogy} and Table \ref{tab:fm-pair} can explain this result. The high recall of the popularity algorithm on the Frappe dataset exhibits a high tendency on popular items in this dataset. On the other hand the presence of bias parameters $w_i$ and $w_j$ in the FM-Pair model can learn such biases, that turns to be very effective on training the model in popularity-skewed datasets such as Frappe. The effectiveness of such bias parameters on CF models have also been shown in previous studies (e.g. ~\cite{Koren09}).

On the two datasets of Amazon, as you can see in Figure \ref{fig:converge-cd}, the FM-Pair-All converge faster, most likely due to the larger number of training samples, but fails to reach the performance of FM-Pair and FM-Pair-CD. The FM-Pair-CD performs better that the other two methods even with smaller number of epochs and thus it is an effective model to leverage cross-domain auxiliary feature.

\begin{figure}[t]
	\centering
	\includegraphics[width=\textwidth]{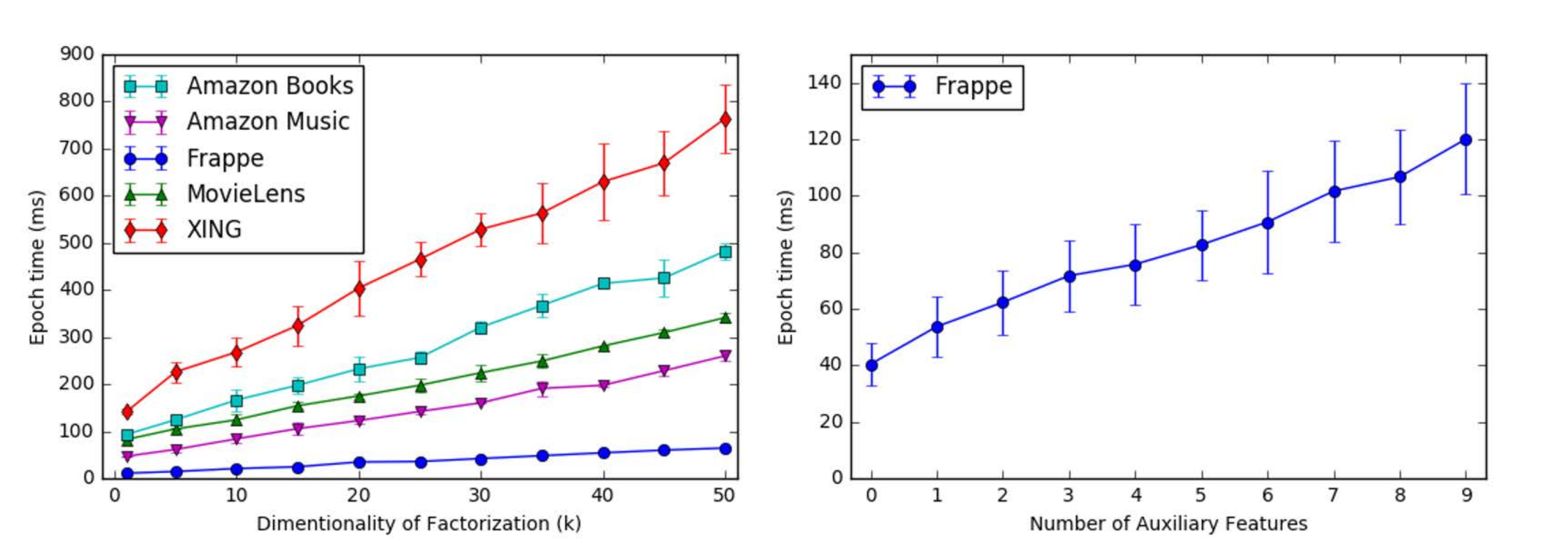}
	\caption{Empirical comparison of the the training time of different datasets based on the dimensionality of factorization (left chart) and number of auxiliary features (right chart).}
	\label{fig:time}
\end{figure}

In Section \ref{sec:complexity} we showed that the complexity of FM-Pair is linear in dimensionality of factorization (parameter $k$) and the number of auxiliary features ($|\mathbf{z}|$). Experimental results also confirms the linearity of FM-Pair. Figure \ref{fig:time} illustrates the influence of the two parameters $k$ (left chart) and $|\mathbf{z}|$ (right chart) on the epoch time of different datasets (the effect of parameter $|\mathbf{z}|$ is only illustrated on the Frappe datasets since this is the only dataset with \textit{multiple} context features). The reported epoch time is the average epoch time on four-fold cross-validation experiments and the bars indicate the standard deviation of the four folds. As you can see in the two charts for both parameters the epoch time grows linearly (with small errors) and thus the linearity of FM-Pair can be confirmed.

\subsection{Using WrapRec}
The implementation of this work is published in the WrapRec toolkit. WrapRec is an  open source evaluation framework for recommender systems that can wrap algorithms from different frameworks and evaluate them under same setup. WrapRec is written is C\# and can be used in multiple platforms. WrapRec can be used as a command-line tool. To use WrapRec all setting need to be defined in one configuration file. The configuration file specifies the model and its parameters, how the dataset should be read and split, and how the evaluation should be done.
The command-line tool can be downloaded from the WrapRec website\footnote{\url{http://wraprec.crowdrec.eu/}} and can be simply used as:
\begin{itemize}
\item (Windows): \texttt{WrapRec.exe [path-to-config-file]}

\item (Linux and Mac):  \texttt{mono WrapRec.exe [path-to-config-file]}
\end{itemize}

Details about the format of the configuration file and usage of WrapRec can be found in the WrapRec website. The experiments on this paper can be reproduced by using the configuration file that is defined for the experiments\footnote{The configuration file will be release in the WrapRec website.}. 

\section{Conclusion and Future Work}
In this work we introduce FM-Pair, an adaptation of Factorization Machines with a pairwise learning-to-rank optimization technique. In contrast to the original model of Factorization Machines, FM-Pair can be used effectively for datasets with implicit feedback, thus addressing a wider range of problems in recommender systems. In this work we show that for ranking problems, FM-Pair is more effective than the standard FMs even on datasets with explicit feedback. FM-Pair leverages a pairwise learning-to-rank method inspired by the Bayesian Personalized Ranking (BPR) criterion, which optimizes the model parameters for ranking. Similar to the standard Factorization Machines, FM-Pair can exploit additional features such as context, user and item attributes, cross-domain information and any discrete or real-valued auxiliary features. In this work we also propose to apply FM-Pair for context-aware and cross-domain collaborative filtering problems. 
%We performed several experiments to show the efficiency and effectiveness of FM-Pair on learning from user feedback.
%When there are no auxiliary features, FM-Pair is performing at least as good as the state-of-the-art, if not better. However, FM-Pair shines when there are auxiliary features in the dataset. FM-Pair can exploit additional features easily and effectively and results to better recommendations while the computational complexity remains linear. Experimental results on context
Experimental results on four datasets with implicit or explicit feedback showed the effectiveness of FM-Pair for datasets with implicit or explicit feedback with or without auxiliary features. We showed that when no auxiliary features are exploited FM-Pair is at least as accurate as state-of-the-art methods such as BPR-MF, if not more. However, FM-Pair shines with its ability to easily exploit additional features without any effort to adapt the underlying model. For the two task of context-aware and cross-domain CF we show that FM-Pair is effective on exploiting such features. The model can be trained without much of overhead on training time while considerable improvement can be achieved by exploiting additional features. Comparison of FM-Pair with GPFM, which is also capable of exploiting context features, exhibits superiority of FM-Pair in terms of accuracy and complexity.

In this work we also observed that the trivial implicit-to-explicit mapping is not an effective way of using FMs for learning-to-rank from datasets with implicit feedback. In fact, the standard FMs are not optimized for ranking and even for datasets with explicit feedback, standard FMs are not effective for ranking problems.

We also analyzed the convergence and complexity of FM-Pair in the tested datasets. An interesting observation was the ability of FM-Pair to leverage item biases that turns to be very effective for the Frappe dataset, which is a popularity-skewed dataset.
We also empirically show that FM-Pair scales linearly on dimensionality of factorization and number of features.

As a future work, the proposed methods for context-aware and cross-domain CF can be further investigated by studying the effect of selected features on the performance of the recommendations. For example, for the task of cross-domain CF, as we mentioned earlier, the features from source domain can be transferred with several possibilities. In this work the features correspond to items in the source domains. The number and characteristics of the selected items in source domains is subject to further studies. Similarly, for the task of context-aware recommendation, the contribution of different context features can be adjusted by feature engineering.

%Furthermore, the optimization model can go beyond pairwise methods and techniques like list-wise learning-to-rank can be exploited as well. 
In this work we adapted a pairwise optimization technique for Factorization Machines. Further studies can be done to apply other optimization techniques such as list-wise learning-to-rank methods for Factorization Machines.

% Acknowledgments
\begin{acks}
This work is supported by funding from EU FP7 project under grant agreements no. 610594 (CrowdRec)
\end{acks}

% Bibliography
\bibliographystyle{ACM-Reference-Format-Journals}
\bibliography{ref}
                             % Sample .bib file with references that match those in
                             % the 'Specifications Document (V1.5)' as well containing
                             % 'legacy' bibs and bibs with 'alternate codings'.
                             % Gerry Murray - March 2012

% History dates
%\received{February 2007}{March 2009}{June 2009}

% Electronic Appendix
%\elecappendix

%\medskip

%\section{This is an example of Appendix section head}

\end{document}